\documentclass[12pt]{article}
\usepackage{pic03}
\usepackage{hyperref}
\usepackage{url}
\usepackage{graphicx}

\newcommand{\mvud}{V_{ud}}
\newcommand{\mvus}{V_{us}}
\newcommand{\mvub}{V_{ub}}
\newcommand{\mvcd}{V_{cd}}
\newcommand{\mvcs}{V_{cs}}
\newcommand{\mvcb}{V_{cb}}
\newcommand{\mvtd}{V_{td}}
\newcommand{\mvts}{V_{ts}}
\newcommand{\mvtb}{V_{tb}}
\newcommand{\tvud}{$|V_{ud}|$}
\newcommand{\tvus}{$|V_{us}|$}
\newcommand{\tvub}{$|V_{ub}|$}

\newcommand{\tvcb}{$|V_{cb}|$}
\newcommand{\tvtd}{$|V_{td}|$}
\newcommand{\tvts}{$|V_{ts}|$}

\begin{document}

\title{\bf MEASUREMENT OF CKM ELEMENTS AND THE UNITARITY TRIANGLE}
\author{Bo\v stjan Golob\\
{\em Faculty of Mathematics and Physics, University of Ljubljana,}\\
{\em Jadranska 19, 1000 Ljubljana, Slovenia\footnotemark }}

\maketitle
\renewcommand{\thefootnote}{\fnsymbol{footnote}}
\setcounter{footnote}{1}
\footnotetext{Representing the Belle
Collaboration; also with the Jo\v zef Stefan Institute.}
\renewcommand{\thefootnote}{\arabic{footnote}}
\setcounter{footnote}{0}
%
%
%
%
%
%
\vspace{4.5cm}
%

\baselineskip=14.5pt
\begin{abstract}
An overview of experimental determinations of different CKM elements
is given with an emphasis on $|V_{\rm ub}|$ and $|V_{\rm cb}|$
extraction. Measurements are compared to the Standard Model
predictions and constraints on the unitarity triangle, arising from
combination of measurements, are presented.
\end{abstract}
\newpage

\baselineskip=17pt

\section{Introduction}
\label{sec:Intro}
\begin{figure}[htbp]
  \centerline{\hbox{ \hspace{0.2cm}
    \includegraphics[width=7cm]{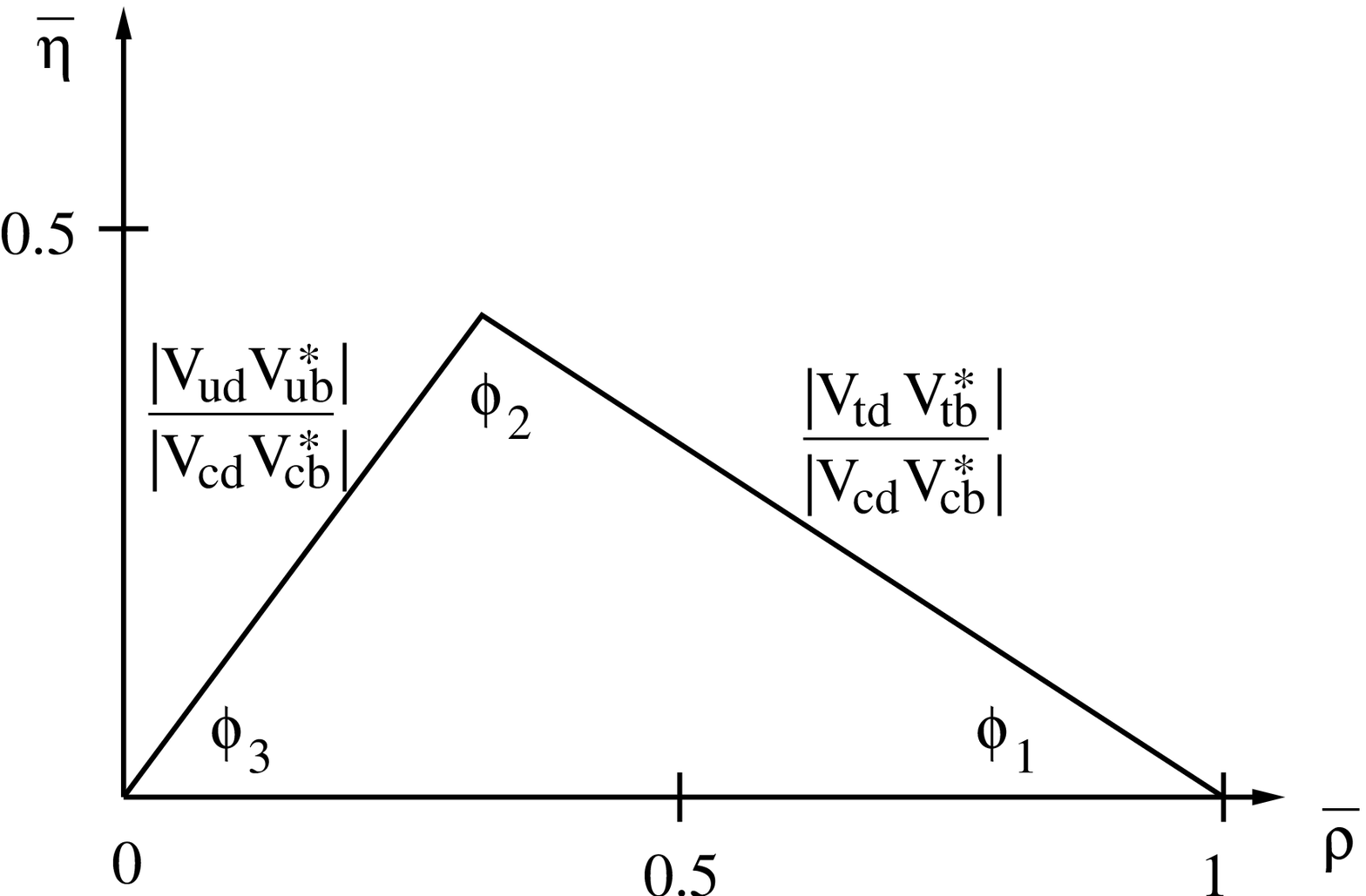}
    \hspace{0.3cm}
    \includegraphics[bb = 73 243 652 678, width=7cm, clip]{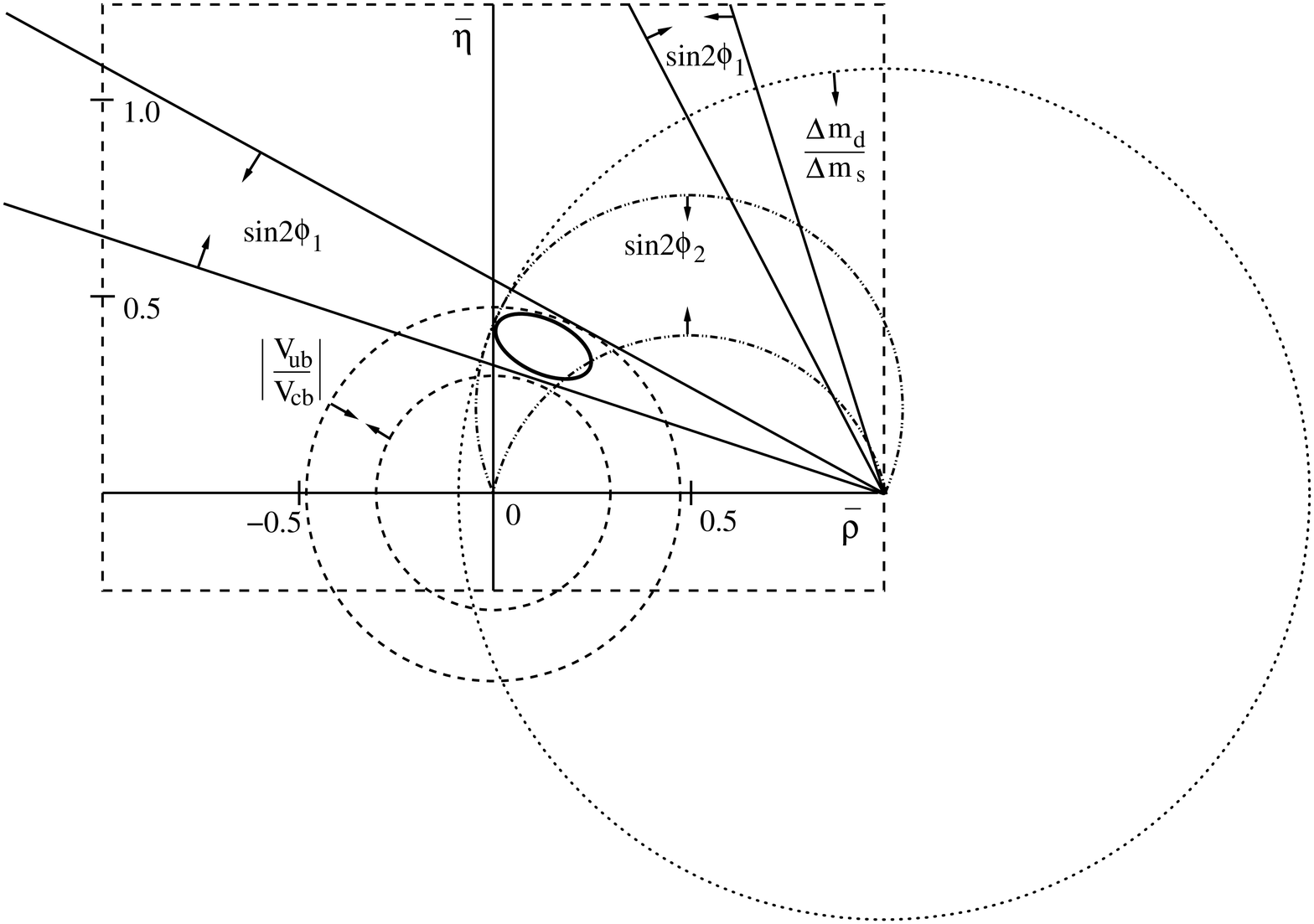}
    }
  }
 \caption{\it Presentation of CKM matrix unitarity in the form of the
unitarity triangle (left). Marked regions illustrate the constraints imposed
on the sides and angles of the triangle by different measurements in
the $B$ meson system (right).
    \label{fig:UT}}
\end{figure}

The general form of the Cabbibo-Kobayashi-Maskawa (CKM) \cite{CKM1} 
matrix can be 
described 
using the generalized Wolfenstein parametrization\footnote{$\overline{\rho}=\rho(1-\lambda^2/2),~
\overline{\eta}=\eta(1-\lambda^2/2)$} \cite{CKM2} as
\begin{equation}
V=\pmatrix{ \mvud & \mvus & \mvub \cr
            \mvcd & \mvcs & \mvcb \cr
            \mvtd & \mvts & \mvtb \cr }
=\pmatrix{ 1-{1\over 2}\lambda^2 & 
\lambda  & A\lambda^3(\rho - i\eta) \cr
-\lambda & 
1-{1\over 2}\lambda^2 &
A\lambda^2  \cr
A\lambda^3(1-\overline{\rho}-i\overline{\eta}) &
-A\lambda^2 &
1 \cr }+{\it O}(\lambda^4)~.
\label{eq:CKMform}
\end{equation}
Product of the first and the last column of the CKM matrix under the
unitarity requirement yields the standard form of the unitarity
triangle (UT), presented in the complex $(\overline{\rho},\overline{\eta})$
plane in Fig. \ref{fig:UT} (left)\footnote{Angles
$\phi_1,~\phi_2,~\phi_3$ are frequently denoted as $\beta,~\alpha$ and
$\gamma$, respectively.}.
Measurements of various observables, from which individual CKM
elements can be determined, impose constraints on the sides and angles
of the UT, shown in Fig. \ref{fig:UT}
(right). The overlap area determines an allowed region for the position of
the UT apex. Comparison of individual measurements, described
below, enable a
consistency check of the CKM description of quark mixing and can give
hints of yet unobserved phenomena. 

\section{\bf{\tvus , \tvud}}
\label{sec:vusvud}
The \tvus~element of the CKM matrix has traditionally been obtained from
the measured branching fractions of $\rm{K}_{\ell 3}$ decays, the
semileptonic decays
of charged and neutral kaons $K\to\pi\ell\nu$. Recently, a new
measurement of the $K^+$ branching fraction in the electron decay mode was
presented by the Brookhaven E865 experiment \cite{E865}. The result
$Br(K^+\to\pi^0e^+\nu)=(5.13\pm 0.02_{\rm stat.}\pm 0.09_{\rm
syst.}\pm 0.04_{\rm norm.})\%$ is about 2.3 standard deviations
higher than the world average of previous measurements of this quantity
\cite{PDG}. The measured branching ratio, where the last error comes
from the uncertainty in the normalization decay modes used in the
experiment, can be converted into the
value of $|\mvus| = 0.2272\pm0.0020_{\rm exp.}\pm 0.0018_{\rm th.}$. The
error on this value receives a significant contribution from
the theoretical estimate of the $K^\pm$ decay form factor \cite{cirigliano}.
Using the value of $|\mvud|=0.9740\pm 0.0005$, evaluated most 
precisely from the super-allowed 
nuclear Fermi
beta decays \cite{towner}, and the above result for \tvus, the
unitarity requirement in the first row of the CKM matrix is perfectly
satisfied\footnote{The value of \tvub~ can be safely neglected.}. 
A possible disagreement between the world averages of 
$Br(K^+\to\pi^0e^+\nu)$ and $Br(K^0\to\pi^-e^+\nu)$\footnote{In the
following the notations include the charge conjugate modes, unless
explicitly stated otherwise.} will be addressed
by the forthcoming precision results from KLOE and NA48 experiments. 

\section{\bf{\tvcb , \tvub}}
\label{sec:vubvcb}
 
Due to a reasonably good theoretical understanding and a 
satisfactory statistical
power of the sample, the 
determination of \tvcb~and \tvub~elements relies on the measurements
of semileptonic $B$ meson decays, using either exclusive or inclusive
methods. Within the exclusive methods, the $B\to
D^\ast\ell\nu$ decay channel is the most prominent one for the $b\to c$
transitions. Decays into
$\pi,~\rho$ and $\omega$ mesons and a lepton pair were used to probe
the $b\to u$ transitions. Inclusive
methods comprise a determination of the total B meson semileptonic
width combined with measurements of various moments of distributions
that reduce the theoretical uncertainties entering the CKM element
extraction. 
 
\begin{figure}[htbp]
  \centerline{\hbox{ \hspace{0.2cm}
    \includegraphics[width=8cm]{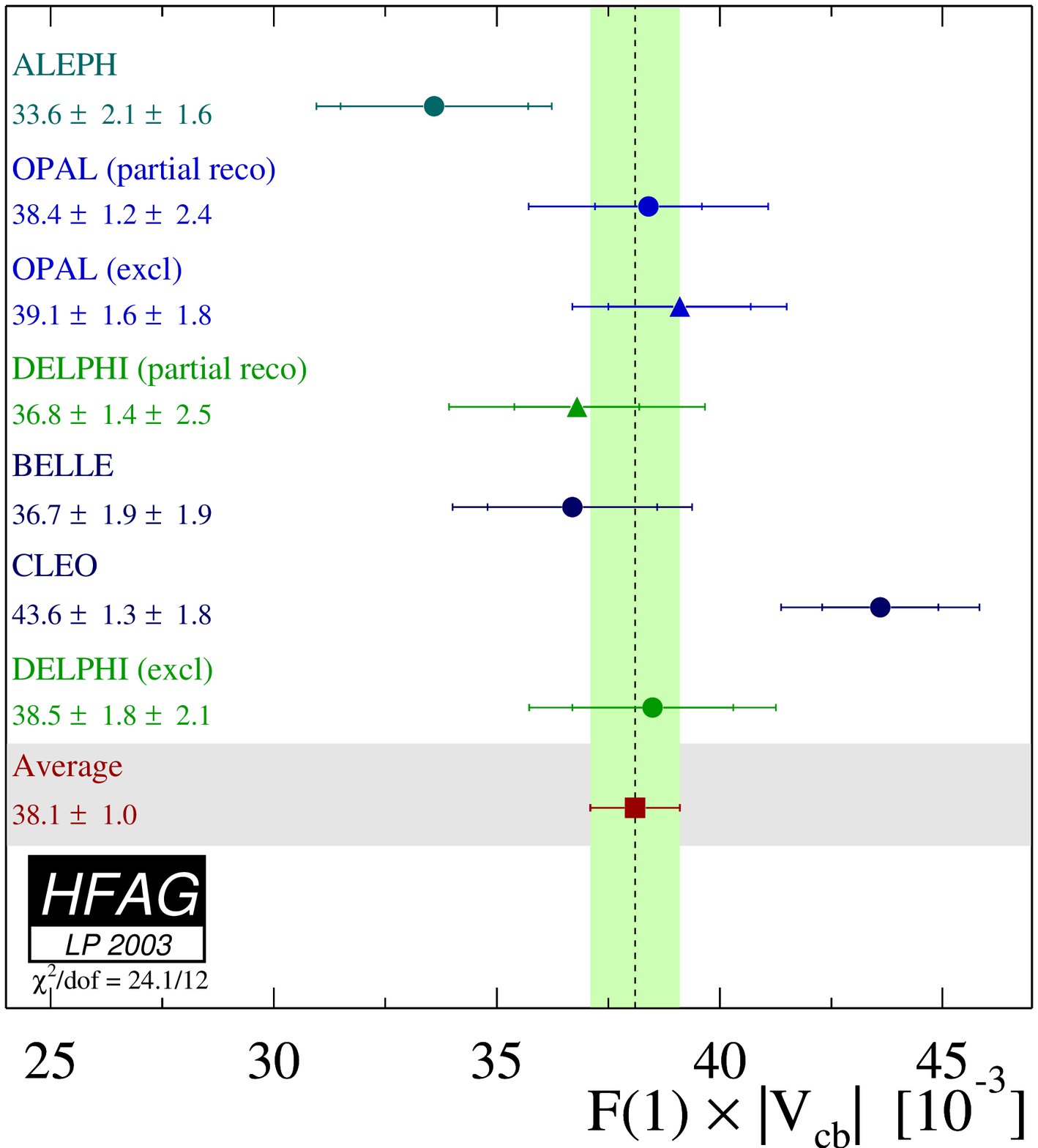}
    \hspace{-1cm}
    \includegraphics[width=8cm]{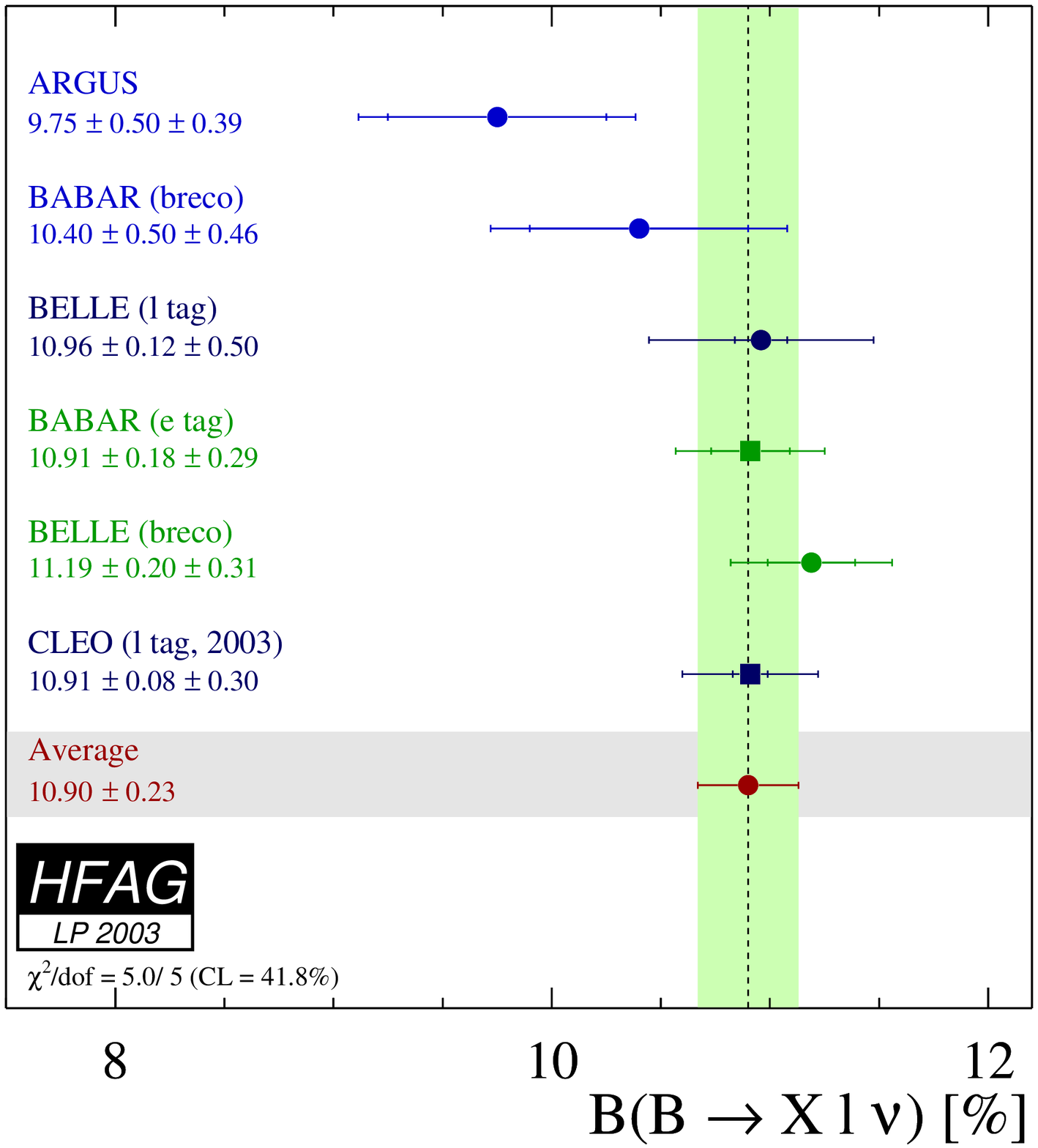}
    }
  }
\caption{\it Summary of \tvcb~ determination using $B\to D^\ast\ell\nu$
(left) and inclusive semileptonic B meson branching fraction (right)
as given by \cite{hfag}.
    \label{fig:vcb_ex_sum}} 
\end{figure}

The differential decay rate for the $B\to
D^\ast\ell\nu$ can be expressed in terms of a four-velocity transfer 
$w=v_Bv_{D^\ast}$ using a single $B$ meson decay form factor
$\it{F}(w)$, which coincides with the Isgur-Wise function up to
the heavy quark symmetry breaking terms \cite{caprini}. Experimentally,
the efficiency
corrected differential decay rate $d\Gamma/dw$ is measured and
extrapolated to the maximum momentum transfer ($w=1$). 
Measurement uncertainties are dominated by the systematic
errors as seen in Fig. \ref{fig:vcb_ex_sum} (left) \cite{hfag}. 
At the $Z^0$
energies, the $w$ resolution is influenced by a spread of $B$ meson
momenta. 
Another important source of the systematic error is the
modeling of the $D^{\ast\ast}$ background. At the $\Upsilon(4S)$
energies the kinematic constraints enable a better $w$ resolution and 
an effective $D^{\ast\ast}$ rejection, but 
the efficiency for the reconstruction of low momentum pions 
from $D^{\ast +}\to D^0\pi^+$ falls rapidly at low values of $w$.
Using the interval of values
${\it F}(1)=0.91\pm 0.04$ \cite{PDG} and the average value of ${\it
F}(1)|\mvcb|$ one obtains 
\begin{equation}
|\mvcb|=(41.9\pm1.1_{exp.}\pm 1.8_{{\it F}(1)})\times 10^{-3}~,
\label{eq:vcb_ex_ave}
\end{equation}
where the first error represents the experimental and the second the
theoretical uncertainty. 

The relation between the semileptonic decay width of $B$ mesons and
the \tvcb~ can be written as 
$Br(b\to c\ell\nu)/\tau_B = {\it K}_{th}|\mvcb|^2$. While the average of
measurements of the left hand side of equation contributes about 1\% to
the relative error on the \tvcb, the theoretical factor 
${\it K}_{th}$ introduces 
an error of about 5\% \cite{PDG}. The solution is provided in the framework
of operator product expansion, where the semileptonic width is
expressed to order 
${\it O}(1/m_b^2)$ using two parameters and heavy quark masses. The same
parameters enter the expressions for differential distributions of
several observables like the energy of the lepton in semileptonic
decays, the invariant mass of the produced hadronic system or the energy of the
photon in radiative B meson decays \cite{luke}. Measurements of moments of those
distributions enable the relevant parameters determination and hence
help reducing the theoretical uncertainty in the extraction of \tvcb~
from the measured semileptonic width. A summary of $Br(B\to c\ell\nu)$
measurements at $\Upsilon(4S)$ energies is shown in
Fig. \ref{fig:vcb_ex_sum} (right) \cite{hfag}. 
The precision is almost identical to
the measurements performed by the LEP experiments, which benefit from
the fact that the total lepton energy spectrum can be used in
analyzes. 
Combination of results
together with the measured $B$ meson lifetime yields a value of
$\Gamma(b\to c\ell\nu)=(0.43\pm 0.01)\times 10^{-10}$~MeV \cite{PDG}. 

Cleo and Delphi experiments measured the moments (up to the third) of 
distributions mentioned above \cite{moments}. 
Taking as an input
the value of the semileptonic width, \tvcb ~evaluations \cite{battaglia}
can be summarized as \cite{stocchi}
\begin{equation}
|\mvcb|=(41.2\pm 0.7_{exp.}\pm 0.6_{th.})\times 10^{-3}~.
\label{eq:vcb_inc_ave}
\end{equation}
The experimental error includes contributions from moments and
semileptonic width 
measurements, while the theoretical error reflects the 
uncertainties due to the perturbative QCD and $1/m_b$ series truncation. 
 
\begin{figure}[htbp]
  \centerline{\hbox{ \hspace{0.2cm}
    \includegraphics[width=6.5cm]{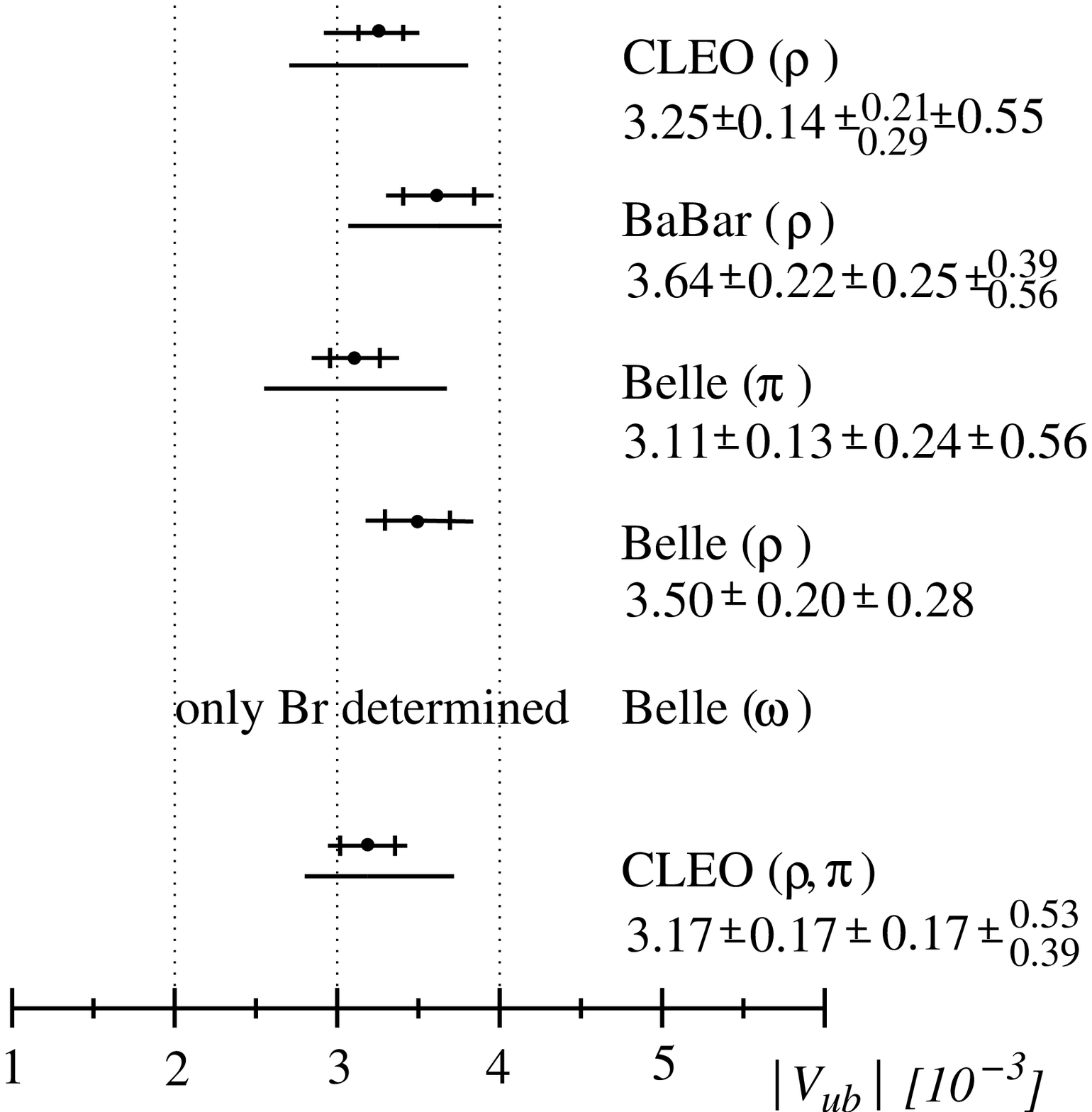}
    \hspace{0.3cm}
    \includegraphics[width=8.5cm]{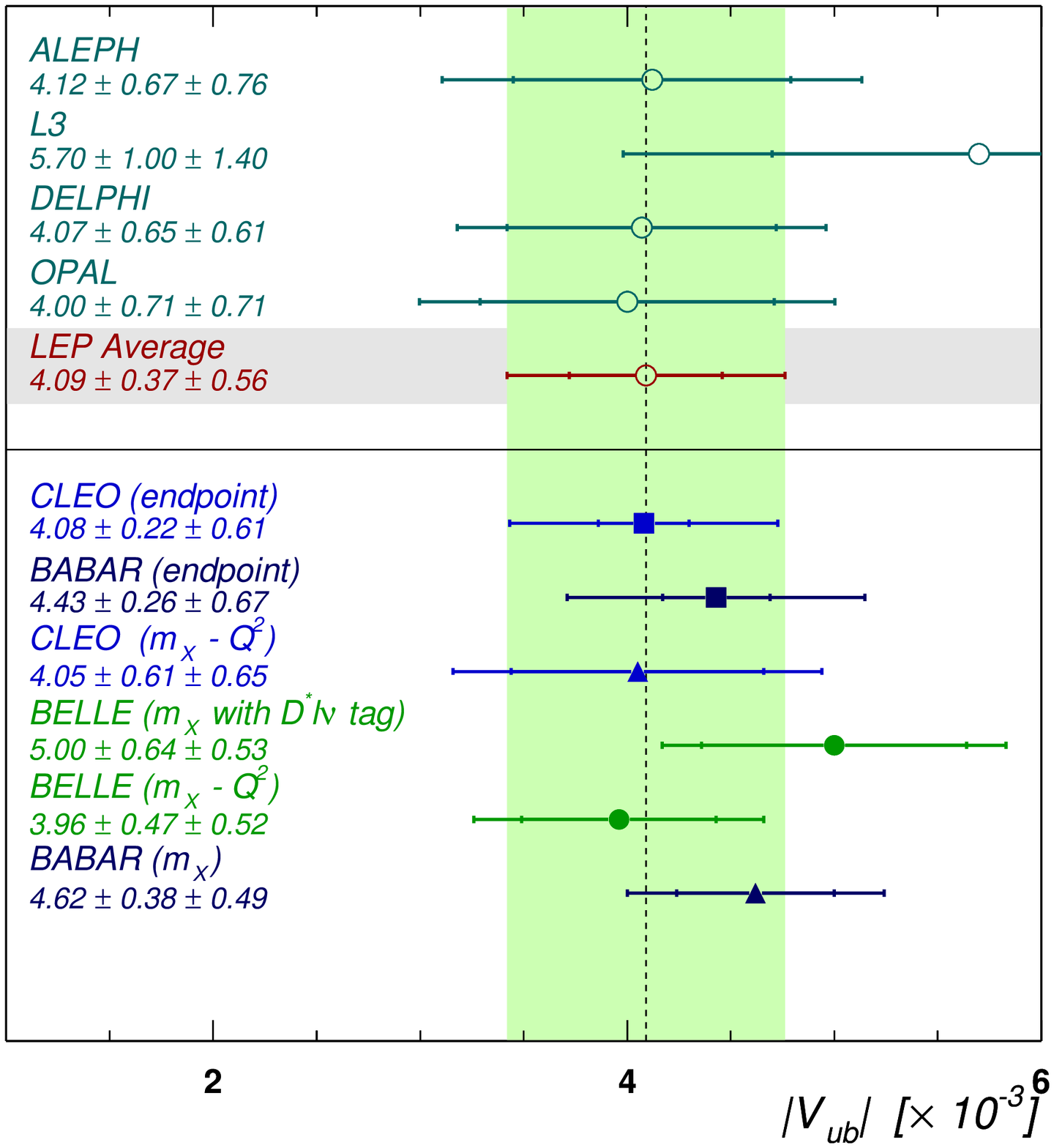}
    }
  }
\caption{\it Summary of \tvub~ determinations using 
$B\to \pi (\rho,~\omega)\ell\nu$
(left) and inclusive $B\to X_u\ell\nu$ branching fraction (right) \cite{hfag}.
    \label{fig:vub_ex_sum}}
\end{figure}

The exclusive methods of \tvub~ determination consist of $Br(B\to \pi (\rho,
\omega) \ell\nu)$ measurements over a limited interval of lepton pair
invariant mass ($q^2$), lepton momentum or lepton energy, in order to
suppress the dominant $b\to c$ background. One or more dimensional
fits to distributions of kinematical variables, separating the $b\to
u$ transitions from the background, make use of isospin relations
between the charged and neutral $\pi$ and $\rho$ decay modes. In order
to obtain the full $Br(B\to X_u\ell\nu)$, where $X_u$ represents a
light meson, an extrapolation to the full range of fitted observables
is necessary, introducing a model dependence. The results are shown in
Fig. \ref{fig:vub_ex_sum} (left). 
Error bars mark the statistical uncertainty, while the total line
represents the quadratic sum of statistical and experimental
systematic errors; for the latter the largest
contributions arise from detector simulation used in neutrino reconstruction
and $b\to u\ell\nu$ background modeling. Lines below the results show
the theoretical uncertainties. They are obtained using
various form factors, calculated from lattice QCD, light cone sum
rules or quark models, with different methods and are thus not
directly comparable. The Belle experiment made a first measurement of
$Br(B \to\omega\ell\nu)$ \cite{belle_omega}, 
however the value of \tvub~is not quoted. In
order to reduce the theoretical uncertainty, the Cleo experiment has
performed a measurement of $Br(B\to\pi~(\rho) \ell\nu)$ (last line of
Fig. \ref{fig:vub_ex_sum} (left)) in distinct $q^2$ bins \cite{cleo_q2}. The
advantage of such an approach is a lower systematic error arising from
the $q^2$ dependence of the efficiency, and the discriminating power of
the $q^2$
distribution for different form factor models. For the exclusive \tvub~
determination no official average is provided by the Heavy Flavor
Averaging Group \cite{hfag} up to date. Using the first four results
of Fig. \ref{fig:vub_ex_sum} (left), and treating half of the systematic
error as completely correlated (due to $b\to u\ell\nu$ background), one
obtains the value
\begin{equation}
|\mvub|=(3.32\pm 0.22_{exp.}\pm 0.54_{th.})\times 10^{-3}~,
\label{eq:vub_ex_ave}
\end{equation}
illustrating the uncertainty dominated by the theory. 

The $b\to u\ell\nu$ semileptonic width is sensitive to \tvub~ in an
analog way as the $b\to c\ell\nu$ is sensitive to \tvcb. However, in
isolating the $b\to u$ transitions, one is restricted to a limited
interval of kinematic variables, separating the signal from the major
$b\to c$ background. While stringent requirements on the lepton
energy, hadronic or lepton pair invariant mass in semileptonic decays
lead to a better purity of the selected sample, the selection of
events in a vicinity of the phase
space boundaries introduces large non-perturbative corrections to
theoretical predictions and hence a larger theoretical error \cite{luke2}. 
A number of measurements using individual or several of the above 
mentioned observables for $b\to u\ell\nu$ separation were reported.  
For the reconstruction of the invariant mass of the hadronic system
($m_X$), produced in a semileptonic $B$ meson decay at $B$ factories, 
one needs to
separate the decay products of individual $B$ mesons. A large number of
recorded $B$ decays enables the use of a sample, where
one of the $B$ mesons is fully reconstructed through its $b\to c$
decay modes. Using this method, BaBar collaboration determines the
yield of $B\to X_u\ell\nu$ events at low values of $m_X$ \cite{mx_babar}. The
corresponding \tvub ~ value is shown in the last line of 
Fig. \ref{fig:vub_ex_sum} (right). To separate decay products, 
Belle collaboration uses a
"simulated annealing" technique, based on maximization of a likelihood
function for
a correct and wrong assignment of particles, 
by iteratively exchanging particles
between the two $B$ mesons \cite{mx_belle}. Selection based on 
the reconstructed value of $m_X$ and
$q^2$ yields the result marked in Fig. \ref{fig:vub_ex_sum} (right) as
"Belle ($m_x-Q^2$)". 
Similarly as in the case of the exclusive \tvub~
determination, uncertainties of all results shown are dominated by the
theoretical error introduced by the extrapolation of $b\to u\ell\nu$ yield
to the full interval of kinematic observables. As an illustration, the
average of BaBar, Belle and Cleo measurements \cite{lepton_end} 
using the lepton energy 
for $b\to u$/$b\to c$
separation, yields
\begin{equation}
|\mvub|=(4.15\pm 0.18_{exp.}\pm 0.61_{th.})\times 10^{-3}~,
\label{eq:vub_inc_ave}
\end{equation}
when the theoretical error is taken as completely correlated, while
experimental systematic error is treated as uncorrelated. 

\section{\bf{\tvtd , \tvts}}
\label{sec:vtdvts}

\begin{figure}[htbp]
  \centerline{\hbox{ \hspace{0.2cm}
    \includegraphics[width=8.5cm]{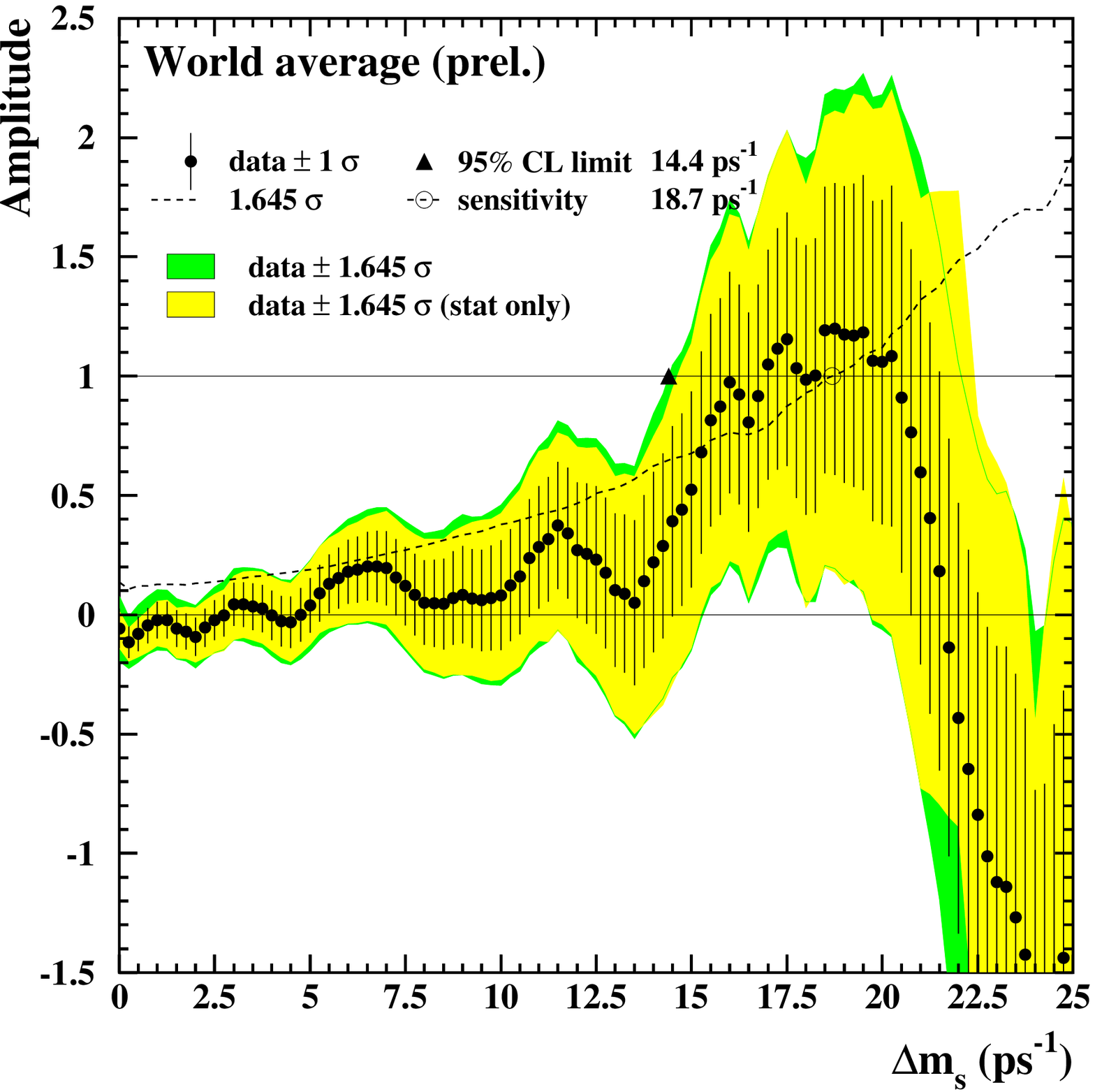}
    \hspace{-1cm}
    \includegraphics[width=6.8cm]{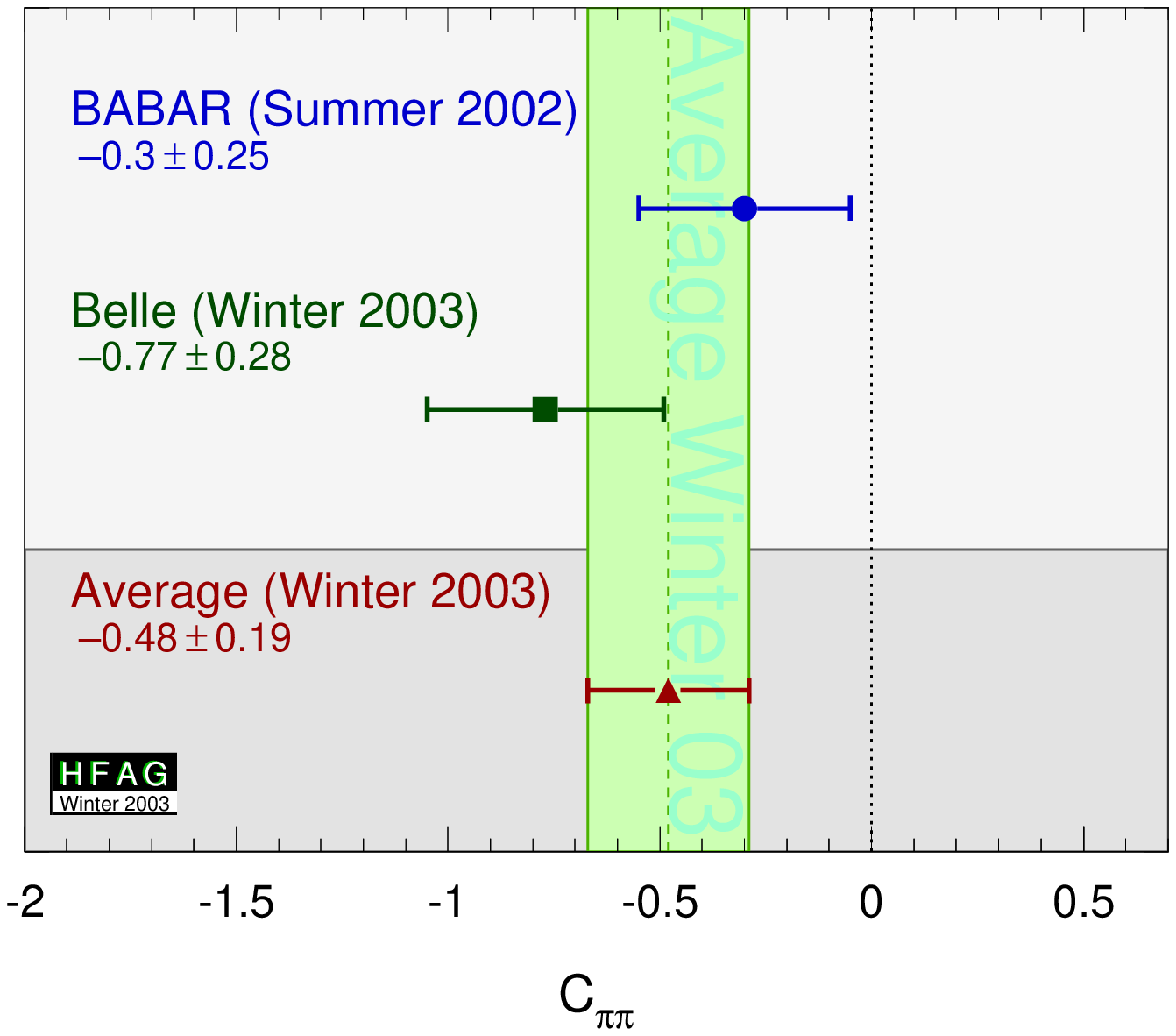}
    }
  }
\caption{\it Summary of $\Delta m_s$ measurements using the amplitude
method (left)\cite{hfag}. Direct CP violation component in
$B^0\to\pi^+\pi^-$ (right)\cite{hfag}.
    \label{fig:delta_ms_sum}}
\end{figure}

The $|\mvtd/\mvts|$ value is determined from the ratio of $B_d$
and $B_s$ meson oscillation frequencies, $\Delta m_d$ and $\Delta
m_s$. The use of the ratio instead of the individual measurements is 
motivated by a partial cancellation of the theoretical uncertainties. 
While the world average of $\Delta m_d$ has a relative error of
1.2\% \cite{hfag} 
and is completely dominated by the measurements performed at 
B factories \cite{Abe},
the $\Delta m_s$ limits steam from LEP experiments and will until the
LHC era remain in the domain of the Tevatron collider. The amplitude
method for the $B_s$ oscillations measurements consists of fitting the
observed decay time distribution with the amplitude of oscillations as a
free parameter. The fitted amplitudes at given values of $\Delta
m_s$ are then converted into the lower limit on the oscillation
frequency. The world average of these measurements is shown in
Fig. \ref{fig:delta_ms_sum} (left), yielding $\Delta
m_s > 14.4~{\rm ps}^{-1}$ at 95\% C.L. \cite{hfag}.  
Taking the $\Delta m_d/\Delta m_s$ ratio and the lattice QCD results
for $\sqrt{B_{B_s}}f_{B_s}/\sqrt{B_{B_d}}f_{B_d}=1.18\pm 0.04\pm{0.12\atop
0.00}$ \cite{lellouch}, one arrives at the limit
\begin{equation}
|{\mvtd \over \mvts}|<0.24~~{\rm at}~95\%~{\rm C.L.}
\label{eq:dms_ave}
\end{equation}

\section{\bf{Angles}}
\label{sec:angles}

A detailed overview of $\sin{2\phi_1}$ measurements, most prominently
in the $B^0\to J/\Psi K_s$ decays, is given in \cite{Abe}. Averaging
results of BaBar and Belle collaborations in all charmonium decay modes
yields $\sin{2\phi_1}=0.736\pm 0.049$ \cite{browder}, a 7\% relative precission
result, where the error is still limited by statistics. An interesting
feature evolves in the measurements of the CP violation using the
$B^0\to\phi K_s$ decay mode with only penguin processes contributing to
the amplitude. These theoretically clean decays are expected to
yield a value of $\sin{2\phi_1}$ consistent with the above quoted
result. However, the present situation reveals a discrepancy of the
two values, the latter being $\sin{2\phi_1}=-0.15\pm
0.33$ \cite{browder}. With an
increased statistics one might expect these measurements to reveal 
new phenomena beyond the SM.  

The angle $\phi_2$ of the UT can be determined examining the
$B^0\to\pi^+\pi^-$ decays \cite{Abe}. The amplitude for 
the process involves a
tree digram as well as a 
significant contribution from the penguin processes, and hence the
interpretation of the CP asymmetry measurements is complicated. While
the Belle result \cite{belle_phi2} 
points to a direct CP violation in these processes 
(Fig. \ref{fig:delta_ms_sum} (right)), the result of the BaBar
collaboration \cite{babar_phi2} is consistent with no direct CP violation.  
Limits obtained on the angle $\phi_2$
from the Belle measurement, taking into account a range
of predictions for the ratio of penguin and tree amplitude
contributions, are $78^\circ < \phi_2 < 152^\circ$ at 95.5\%
C.L.

\section{\bf{UT Constraints}}
\label{sec:constraints}

\begin{figure}[htbp]
  \centerline{\hbox{ \hspace{0.2cm}
    \includegraphics[width=8cm]{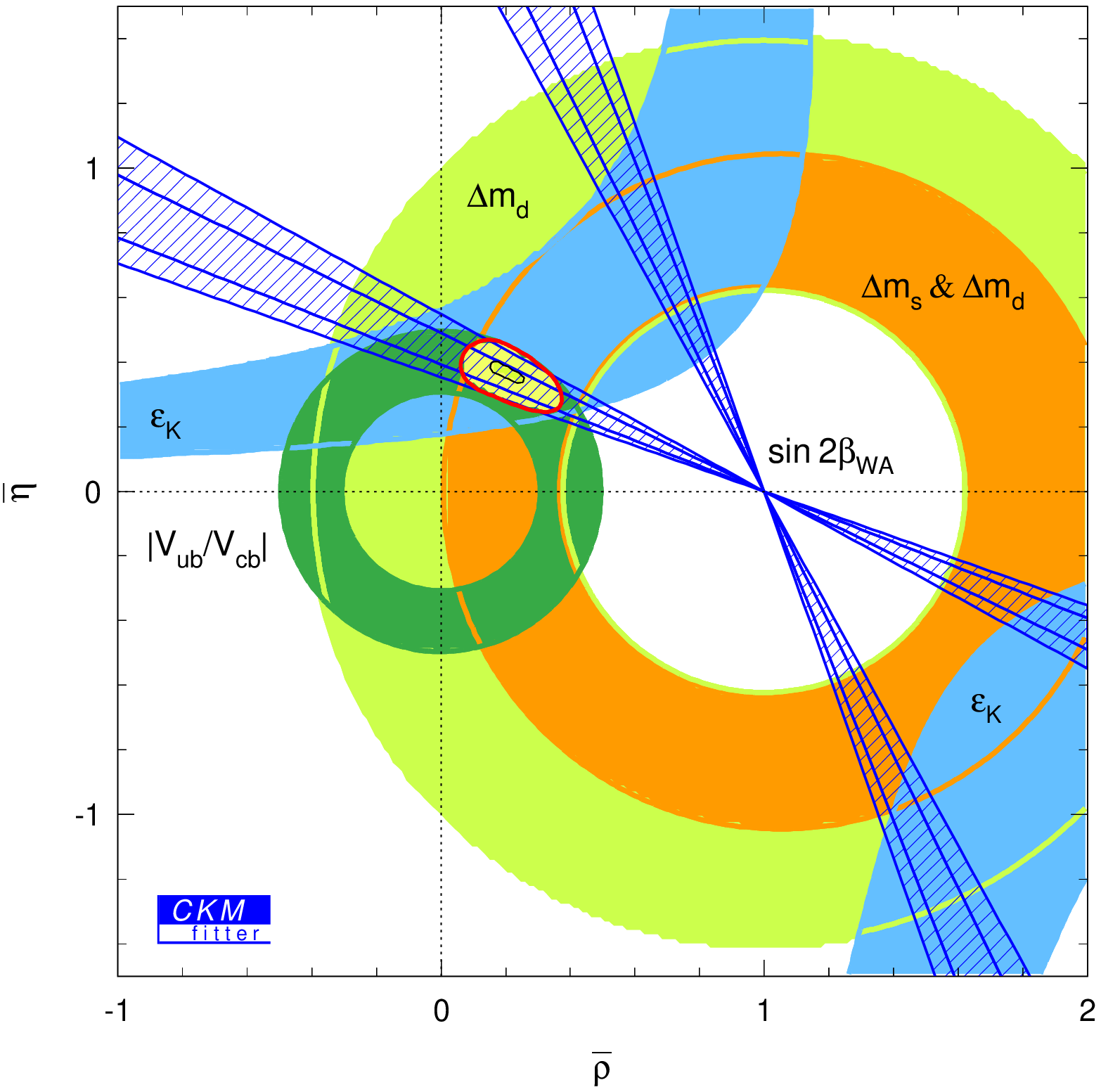}
    \includegraphics[width=6.8cm]{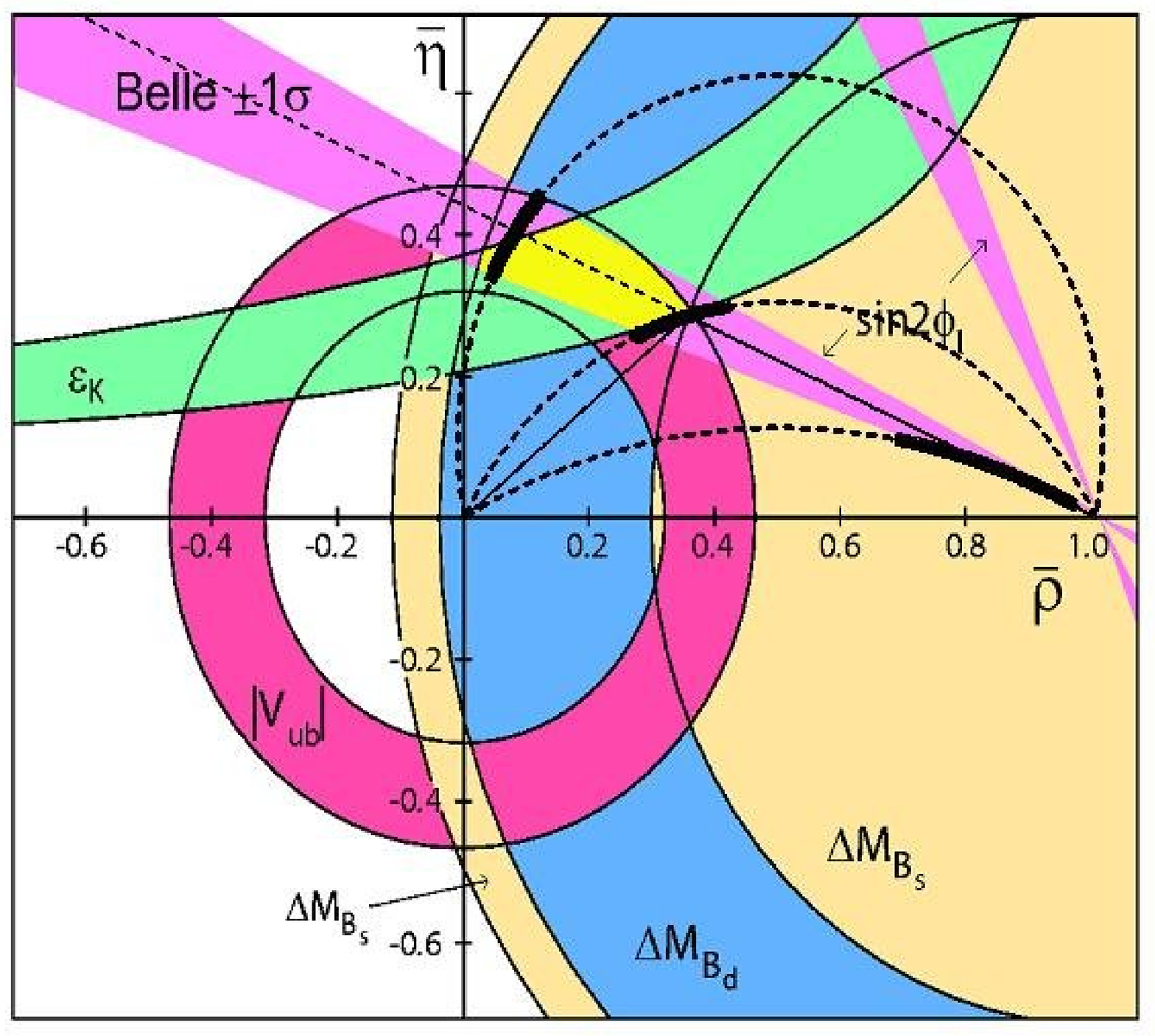}
    }
  }
  \caption{\it Constraint on the position of UT apex in the
$({\overline\rho},{\overline\eta})$ plane obtained from the combined
fit to CKM elements measurements \cite{ckm_fitter}
(left). Illustration of constraints imposed by $\sin{2\phi_2}$
measurement (right) \cite{sagawa}.
    \label{fig:ckm_fit}}
\end{figure}

The CKM Fitter group \cite{ckm_fitter} provides a fit to 
the measured observables and range of theoretical parameters in order 
to constrain the region in the
$({\overline\rho},{\overline\eta})$ plane. In the so called Rfit
approach one maximizes the ${\cal L}(y_{th})={\cal
L}_{exp}(x_{exp}-x_{th}(y_{th}))\cdot{\cal L}_{th}(y_{th})$, where
measurements $x_{exp}$ and theoretical predictions $x_{th}$, depending
on parameters $y_{th}$, enter the experimental part ${\cal
L}_{exp}$. Theoretical part ${\cal L}_{th}$ equals unity if the set of
parameters is within an allowed range of predictions and vanishes
otherwise. Using the set of inputs described at \cite{ckm_fitter} one
obtains a 90\% C.L. region in $({\overline\rho},{\overline\eta})$
plane shown in Fig. \ref{fig:ckm_fit} (left). 
At the present, measurements of $\sin{2\phi_1}$, \tvub~
and $\Delta m_s$ all impose severe constraints on the position of the
UT apex. 
Fig. \ref{fig:ckm_fit} (right) includes results on $\sin{2\phi_1}$
and $\sin{2\phi_2}$ by the Belle experiment \cite{sagawa}, 
showing a consistency of measurements.

\section{\bf{Summary}}
\label{sec:summary}
 
\begin{itemize}
\item{\tvud} is measured with a relative precision of $5\times
10^{-4}$ in the nuclear super-allowed Fermi transitions. The measurement
uncertainty is completely theoretically dominated. 
\item{\tvus}: a new measurement of $K_{e3}$ branching fraction by E865
experiment resolves
the question of the unitarity in the first row of CKM matrix. A
possible disagreement between the charged and neutral $K_{e3}$ decay
modes will be addressed in the future by KLOE and NA48 experiments. 
\item{\tvcb}, determined from $B\to D^\ast\ell\nu$ decays, is limited by
the uncertainty in the decay form factor normalization. Inclusive
determinations will be improved with new measurements of moments of
$E_\ell$, $q^2$ and $m_X$ distributions.
\item{\tvub}: the $q^2$ dependent
measurements have started in exclusive decay channels. 
New inclusive measurements of $Br(b \to
u\ell\nu)$ using $m_X$ and $q^2$ variables will be performed by BaBar and 
Belle experiments. Theoretical ambiguities are expected to be resolved
through tests of different models (in exclusive channels) and determination
of moments of differential distributions ($b\to s\gamma$
transitions). 
\item{\tvtd,~\tvts}: while results on $\Delta m_d$ are already very
precise, further important constraints on UT are hoped for from
$\Delta m_s$ measurements by D0 and CDF experiments.
\item{$\sin{2\phi_1}$} has become a precision measurement, new
phenomena might arise in the CP asymmetry measurements using the $B^0\to \phi
K_s$ decay channel. 
\item{$\sin{2\phi_2}$} determinations have just started and although 
complicated, these measurements will in the future provide an
interesting constraint on the UT.
\end{itemize}


\begin{thebibliography}{99}
\bibitem{CKM1}M.~Kobayashi, T.~Maskawa, Prog. Theor. Phys. {\bf 49},
652 (1973).
\bibitem{CKM2}L.~Wolfenstein, Phys. Rev. Lett. {\bf 51}, 1945 (1983); 
A.J.~Buras {\it et al.}, Phys. Rev. D{\bf 50}, 3433 (1994);
M.~Schmidtler, K.R.~Schubert, Z. Phys. C{\bf 53}, 347 (1992).
\bibitem{E865}A.~Sher {\it et al.}, hep-ex/0305042 (2003).
\bibitem{PDG}K.~Hagiwara {\it et al.}, Phys. Rev. D{\bf 66}, 010001 (2002).  
\bibitem{cirigliano}V.~Cirigliano {\it et al.}, Eur. Phys. J. C{\bf
23}, 121 (2002).
\bibitem{towner}J.C.~Hardy, I.S.~Towner, Eur. Phys. J. A{\bf 15}, 223 (2002).
\bibitem{caprini}I. Caprini {\it et al.}, Nucl. Phys. B{\bf 530}, 153 (1998).
\bibitem{hfag}Heavy Flavor Averaging Group, http://www.slac.stanford.edu/xorg/hfag/
\bibitem{luke}A.F. Falk {\it et al.}, Phys. Rev. D{\bf 53}, 2491 (1996).
\bibitem{moments}S. Chen {\it et al.}, Phys. Rev. Lett. {\bf 87},
251807 (2001); D. Cronin-Hennessy {\it et al.}, Phys. Rev. Lett. {\bf
87} 251808 (2001); A.H. Mahmood {\it et al.}, Phys. Rev. D{\bf 67}
072001 (2003); M. Calvi {\it et al.}, hep-ex/0210046 (2002). 
\bibitem{battaglia}M. Battaglia {\it et al.}, Phys. Lett. B{\bf 556},
41 (2003); C.W. Bauer {\it et al.}, Phys. Rev. D{\bf 67}, 054012 (2003).
\bibitem{stocchi}A. Stocchi, Workshop on the CKM Unitarity Triangle,
Durham, UK, 2003.
\bibitem{belle_omega}K. Abe {\it et al.}, hep-ex/0307075 (2003).
\bibitem{cleo_q2}S.B. Athar {\it et al.}, hep-ex/0304019 (2003).
\bibitem{luke2}M. Luke, Workshop on the CKM Unitarity Triangle,
Durham, UK, 2003.
\bibitem{mx_babar}D. del Re, Electroweak Interactions and Unified
Theories, Moriond, France, 2003.
\bibitem{mx_belle}A. Sugiyama, Electroweak Interactions and Unified
Theories, Moriond, France, 2003.
\bibitem{lepton_end}B. Aubert {\it et al.}, hep-ex/0207081 (2002);
K. Abe {\it et al.}, contributed paper to Lepton Photon 2003, Batavia,
USA, 2003; A. Bornheim {\it et al.}, Phys. Rev. Lett. {\bf 88}, 231803
(2002).
\bibitem{Abe}K.~Abe, XXIII Physics in Collision, Zeuthen, Germany,
2003; Y.~Pan, ibido.
\bibitem{lellouch}L. Lellouch, Nucl. Phys. Proc. Suppl. {\bf 117}, 127
(2003).
\bibitem{browder}T. Browder, Lepton Photon 2003, Batavia, USA, 2003.
\bibitem{belle_phi2}K. Abe {\it et al.}, Phys. Rev. D{\bf 68}, 012001
(2003).  
\bibitem{babar_phi2}B. Aubert {\it et al.}, Phys. Rev. Lett. {\bf 89},
281802 (2002). 
\bibitem{ckm_fitter}H. H$\ddot{\rm o}$cker {\it et al.},
Eur. Phys. J. C{\bf 21}, 225 (2001);  http://ckmfitter.in2p3.fr/
\bibitem{sagawa}H. Sagawa, Flavor Physics and CP Violation, Paris,
France (2003).
\end{thebibliography}
\end{document}